\documentclass[sigplan, authorversion=true]{acmart}
\usepackage[english]{babel}
\usepackage[utf8]{inputenc}
\usepackage[T1]{fontenc}
\usepackage{setspace}

\usepackage{graphicx}

\usepackage{verbatim}

\usepackage{enumitem}
\setlist[itemize]{leftmargin=1em}
\setlist[enumerate]{leftmargin=3ex}
\setlist[description]{leftmargin=1em}

\newcommand{\ifDef}{\texttt{\#ifdef}}

\newcommand{\ifDefBlock}{\ifDef-block}

\definecolor{cEclipseRed}{rgb}{0.6,0,0}
\definecolor{cEclipseGreen}{rgb}{0.25,0.5,0.35}
\definecolor{cEclipsePurple}{rgb}{0.5,0,0.35}
\definecolor{cEclipseBlue}{rgb}{0.25,0.35,0.75}

\usepackage{listings}
\lstdefinelanguage{j} {
    language=Java,
    keywordstyle=[2]{\color{gray!70!black}},
    morekeywords=[2]{@Override},
}

\lstdefinelanguage{properties} {
    tabsize=1,
    commentstyle=\color{cEclipseBlue}, 
    morecomment=[l]{\	},
    morecomment=[l][\color{cEclipseGreen}]{\#}
}

\begin{document}
\newcommand{\mainTitle}{KernelHaven -- An Experimentation Workbench for Analyzing Software Product Lines}

\title[KernelHaven -- An Experimentation Workbench for Analyzing SPLs]{KernelHaven -- An Experimentation Workbench for Analyzing Software Product Lines}

\author{Christian Kröher, Sascha El-Sharkawy, Klaus Schmid}
\affiliation{
  \institution{University of Hildesheim, Institute of Computer Science}
  \city{Universitätsplatz 1, 31141 Hildesheim} 
  \state{Germany} 
}
\email{{kroeher, elscha, schmid}@sse.uni-hildesheim.de}

\begin{abstract}
Systematic exploration of hypotheses is a major part of any empirical research. In software engineering, we often produce unique tools for experiments and evaluate them independently on different data sets. In this paper, we present \textit{KernelHaven} as an experimentation workbench supporting a significant number of experiments in the domain of static product line analysis and verification. It addresses the need for extracting information from a variety of artifacts in this domain by means of an open plug-in infrastructure. Available plug-ins encapsulate existing tools, which can now be combined efficiently to yield new analyses. As an experimentation workbench, it provides configuration-based definitions of experiments, their documentation, and technical services, like parallelization and caching. Hence, researchers can abstract from technical details and focus on the algorithmic core of their research problem.

KernelHaven supports different types of analyses, like correctness checks, metrics, etc., in its specific domain. The concepts presented in this paper can also be transferred to support researchers of other software engineering domains. The infrastructure is available under Apache 2.0: \textit{https://github.com/KernelHaven}. The plug-ins are available under their individual licenses.

\textbf{Video:} https://youtu.be/IbNc-H1NoZU
\end{abstract}

%
%




\copyrightyear{2018}
\acmYear{2018}
\setcopyright{rightsretained}
\acmConference[ICSE '18 Companion]{40th International Conference on Software Engineering }{May 27-June 3, 2018}{Gothenburg, Sweden}
\acmDOI{10.1145/3183440.3183480}
\acmISBN{978-1-4503-5663-3/18/05}

\maketitle

\vspace*{-1em}
\section{Introduction}
In Software Engineering (SE), the systematic exploration of hypotheses often requires the development of tools to analyze early assumptions and their evolution towards evaluating a final approach. This way of validating research concepts requires time and effort for developing appropriate tooling \cite{Merali10}. In particular, these tools are typically only designed for very specific experiments, which requires the design and implementation of its unique data extraction mechanism, data model and analysis as well as standard capabilities like caching, parallelization or logging. Further, each experiment requires a specific setup of process steps or combination of tools to produce the desired results. Reproducing these results in turn requires a precise documentation of the experiment and its data, which is not always the case.

In this paper, we present \textit{KernelHaven} as an experimentation workbench addressing these challenges in the domain of static Software Product Line (SPL) analysis and verification. The focus in this domain is on analyzing and verifying variability information \cite{ThuemApelKaestner+14}, which is typically encoded in different types of artifacts to enable their customization and to derive product variants from the SPL \cite{LindenSchmidRommes07}. Similar to the workbench for checking consistency among different software architecture representations \cite{KonersmannGoedicke12}, KernelHaven addresses the resulting need for information extraction from different artifacts in the SPL domain for variability-based analyses of C-preprocessor SPLs, like Linux. Its open plug-in infrastructure enables a rapid setup for a significant number of experiments, like evaluating approaches for correctness checks, metrics, etc. As an experimentation workbench supports the scientific process at large, e.g., rapid variation of experiments, reuse of components in new experiments, etc., plug-ins realize, for example, individual extraction mechanisms or analysis algorithms, which can be combined efficiently for a specific experiment. Further, the infrastructure offers configuration-based definitions of experiments, their documentation, and technical services, like parallelization and caching. Hence, researchers can abstract from technical details and focus on the algorithmic core of their research problem.

Despite the focus of the tool, the general concepts are much broader. Thus, not only SPL researchers benefit from an off-the-shelf experimentation ecosystem but also researchers of other SE domains, who may adapt these concepts for their specific purposes.

\vspace*{-2ex}
\section{KernelHaven Concepts}
\label{sec:Concepts}
\begin{figure}[t] 
	\centering
		\includegraphics[width=\columnwidth,trim={6,8cm 7,1cm 6,5cm 5,1cm},clip]{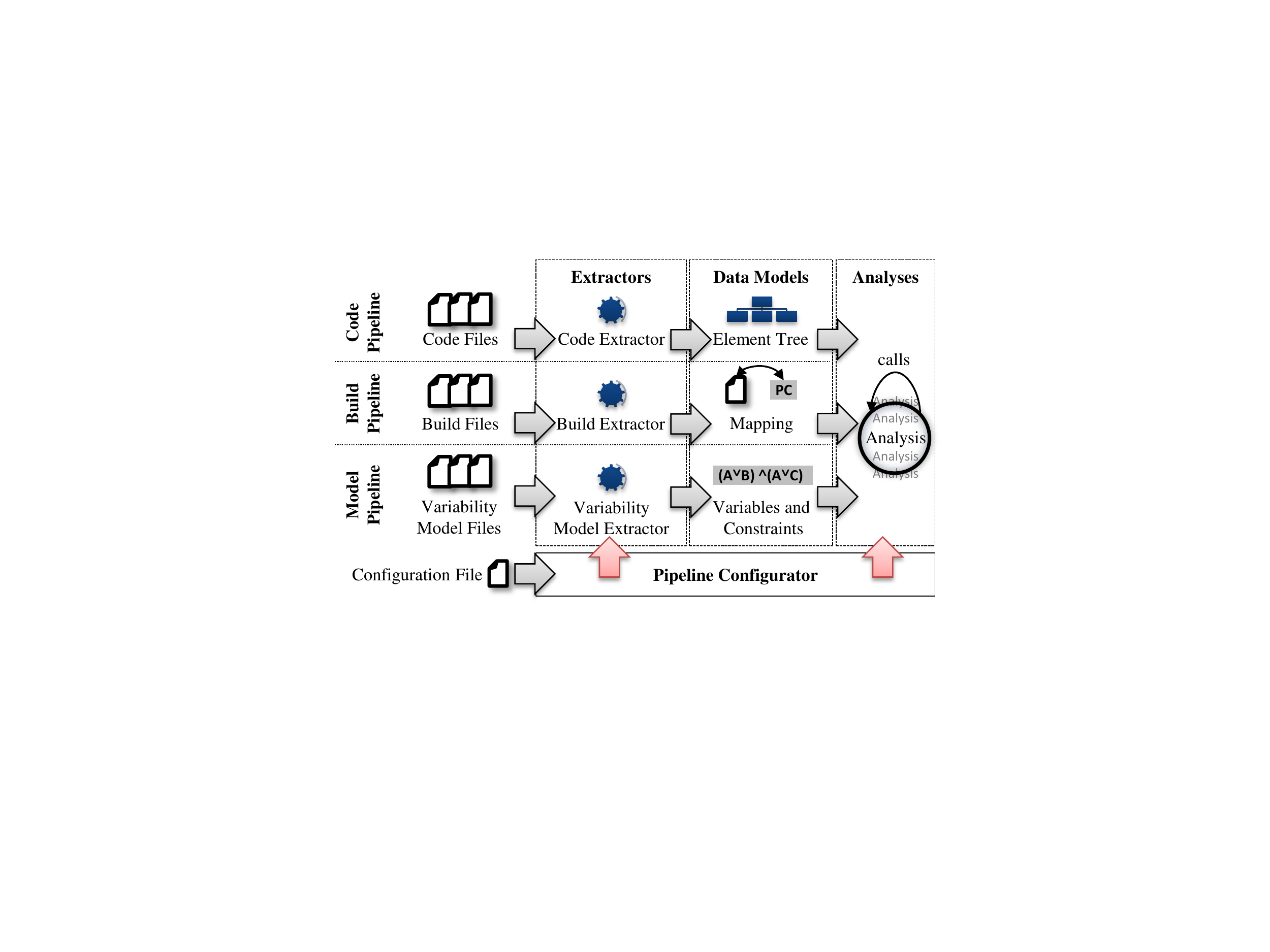}
  \vspace*{-4ex}
  \caption{KernelHaven Architecture}
	\label{fig:Architecture}
  \vspace*{-1ex}
\end{figure}
Variability-based analyses of C-preprocessor SPLs, like Linux, typically consider variability information in code, build, and variability model artifacts. A variability model defines the products of a SPL by individual features and constraints among them \cite{LindenSchmidRommes07}. In code artifacts, \ifDef{-statements} reference these features to make the presence of code blocks dependent on the selection of features of the variability model \cite{SinceroTartlerLohmann+10, KastnerGiarrussoRendel+11}. Similar references in build artifacts control the execution of build rules and the presence of entire files in a product variant \cite{NadiHolt12, NadiBergerKastner+15}. Hence, KernelHaven supplies an analysis with information from these types of artifacts as illustrated in Figure~\ref{fig:Architecture}.

KernelHaven is implemented in Java and consists of three extraction pipelines as well as a pipeline configurator. The upper pipeline in Figure~\ref{fig:Architecture} extracts information from code files using a code extractor. The result of this extraction is an element tree for each of the available code files (we will provide more details on the data models in Section~\ref{sec:Data}). The middle pipeline extracts and provides information from build files as a map of files and their presence conditions (PC in Figure~\ref{fig:Architecture}). These conditions, also called presence implications \cite{DietrichTartlerSchroderPreikschat+12} or make space constraints \cite{NadiHolt12}, define constraints, which must be satisfied to compile and link a specific (set of) file(s). The lower pipeline in Figure~\ref{fig:Architecture} processes variability model files to propositional formulas, which represent the features and constraints defining the planned product variants of the SPL.

The information provided by one or multiple pipelines is input to an analysis. Analyses may call other analyses (cf.\ Section~\ref{sec:Extension}) and consume their results, e.g., for comparison or as input. The infrastructure also provides utility functions to support specific parts of an analysis. For example, it provides conversions from propositional formulas to their conjunctive normal form (CNF) or a solver (SAT4J \cite{Sat4j}), which are typically required in SPL analysis.

The configuration file in the lower left part of Figure~\ref{fig:Architecture} consists of a set of parameters defining the setup of the workbench. The pipeline configurator reads these parameters to configure the extractors and the analysis before their execution. This includes general parameters like input and output locations, the location of available extractors and analyses, and the definition of which extractors and analysis to use. The infrastructure further offers configuration parameters for the features presented in Section~\ref{sec:Features}.

\begin{figure}[t] 
	\centering
		\includegraphics[width=\columnwidth,trim={6,5cm 9,1cm 6,5cm 5,1cm},clip]{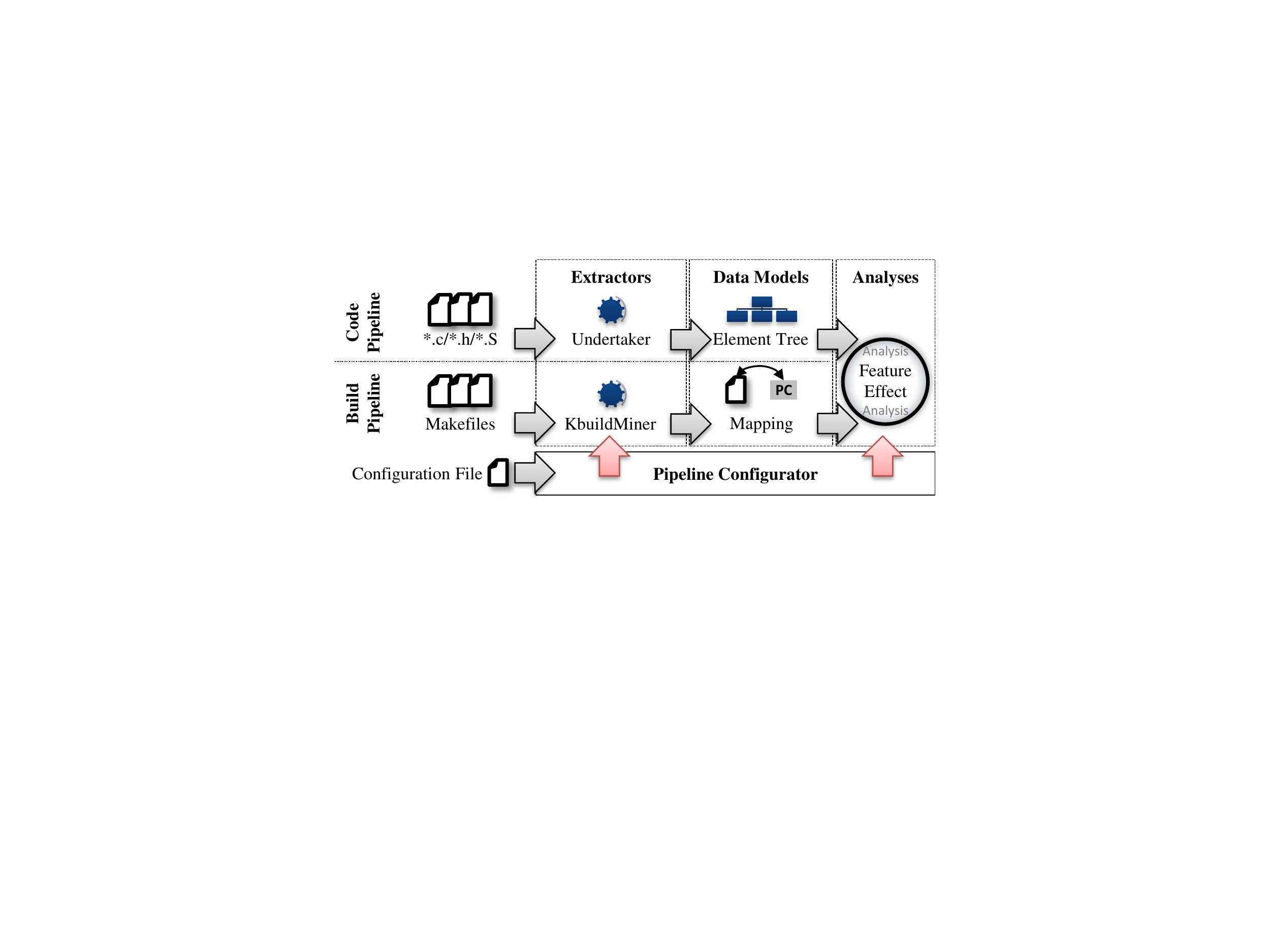}
	\vspace*{-4ex}
  \caption{KernelHaven Instance}
	\label{fig:Instance}
  \vspace*{-1ex}
\end{figure}
The parameter values in a configuration file define a specific instance of KernelHaven as illustrated in Figure~\ref{fig:Instance}. This example is designed to perform a feature effect analysis \cite{NadiBergerKastner+15}, which identifies under which condition the selection of a feature has an impact on (parts of) the variable code and, hence, affects the behavior of the resulting product. As this effect is defined by the presence conditions in the build and code files, the instance in Figure~\ref{fig:Instance} uses the Undertaker \cite{Undertaker} code extractor and the KbuildMiner \cite{KBuildMiner} build extractor for the analysis. While the former extracts \ifDefBlock{s} with their presence conditions, the latter provides such conditions from the build system. Information of the variability model is not needed for this analysis, which results in deleting the respective configuration parameter defining the variability model extractor and the absence of the entire model pipeline in Figure~\ref{fig:Instance}.

\vspace*{-2ex}
\section{KernelHaven Features}
\label{sec:Features}
In this section, we present the core features of KernelHaven, which support researchers in rapid prototyping, conducting similar experiments, as well as reproducing their results. Sections~\ref{sec:Data} and \ref{sec:Extension} discuss features to support flexibly a wide range of experiments. We then present features to reduce the implementation effort for experiment designers in Sections~\ref{sec:Archiving} to \ref{sec:Caching}.

\subsection{Common Data Representation}
\label{sec:Data}
The three extraction pipelines of KernelHaven provide their individual data models, which decouple the data extraction from the analysis and facilitate exchange of alternative extractor realizations. The models contain the relevant information, which is required for conducting the various experiments, independently of the information provided by the diverse extractors. This allows the flexible exchange of alternative extractors, which are based on existing research prototypes designed for different purposes and vary in terms of their advantages.

\begin{figure}[t] 
	\centering
  \includegraphics[trim={5.08cm 11.6cm 4cm 2.75cm},width=\columnwidth,clip]{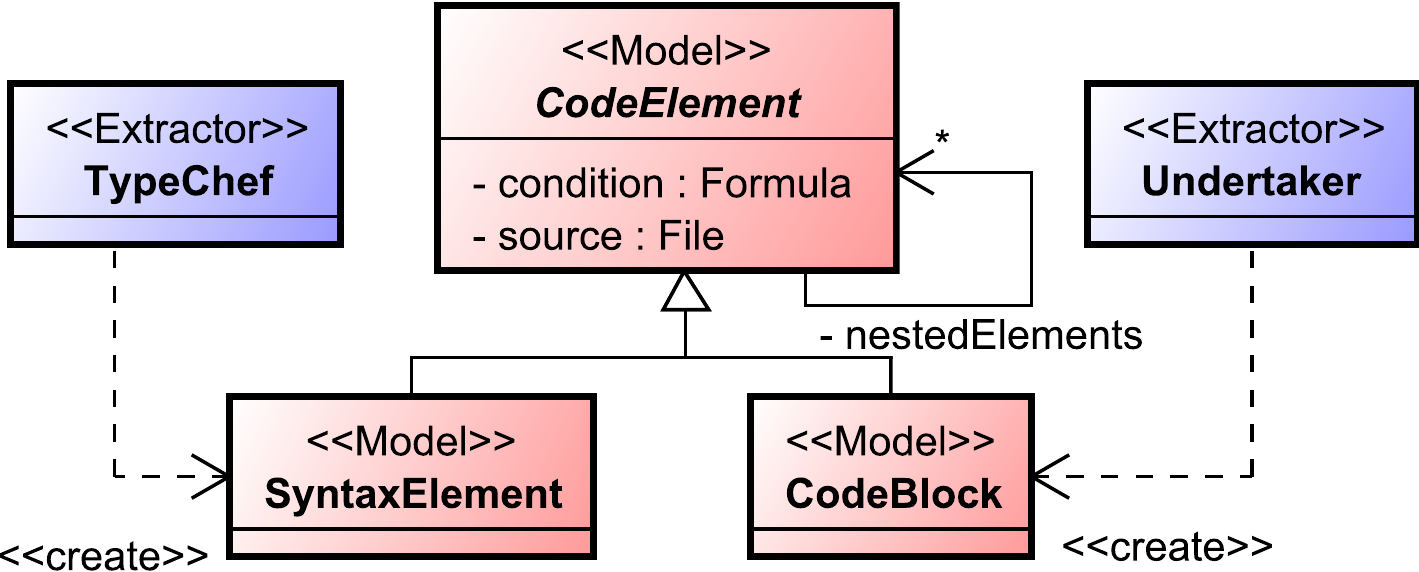}
	\vspace*{-4ex}
  \caption{Excerpt of the Code Data Model in KernelHaven}
	\label{fig:Code Model}
\end{figure}

\begin{figure*}
  \begin{minipage}[t]{0.49\textwidth}
\begin{lstlisting}[language=j]
@Override
protected AnalysisComponent createPipeline() {
  return new FeatureEffectFinder(config,
    new PcFinder(config, getCmComponent(),
      getBmComponent())
  );
}
\end{lstlisting}
  \end{minipage}
  \hspace{1em}
  \begin{minipage}[t]{0.49\textwidth}
\begin{lstlisting}[language=properties]
analysis.class =	ConfiguredPipelineAnalysis
analysis.pipeline =
	FeatureEffectFinder(
	  PcFinder(cmComponent(), bmComponent())
	)
analysis.output.intermediate_results =	PcFinder
analysis.output_writer.class =	ExcelBook
\end{lstlisting}
  \end{minipage}
  \vspace*{-1em}
  \caption{Excerpts of a Programmed and a Configured Analysis Pipeline.}
  \label{fig:Analysis pipeline}
  \vspace*{-3ex}
\end{figure*}

Figure~\ref{fig:Code Model} shows two code extractors and the relevant excerpt of the data model. On the right side of the picture is the Undertaker code extractor, which extracts conditions from \ifDefBlock{s} only. The left side of the picture shows the TypeChef \cite{TypeChef} extractor, which provides an abstract syntax tree (AST) with variability information. Both data structures inherit from \texttt{CodeElement}, which stores the common information. Most analysis implementations use this class as it provides sufficient information. This allows to exchange the fast Undertaker-based extractor with the macro-aware parser of TypeChef, without any modifications of the consuming analysis. However, analyses may operate directly on one of the sub-classes, if they require additional information. For instance, code metrics \cite{El-SharkawyYamagishi-EichlerSchmid17} may operate on \texttt{SyntaxElement}s and, hence, require an extractor generating an AST with variability information, like TypeChef.

The extractors and data models for the other two extraction pipelines are realized in a similar way. The data model for the build information, for example, stores a presence condition for each code file (cf.\ Section~\ref{sec:Concepts}). For the analysis of Linux, we realized an extractor which uses KbuildMiner \cite{KBuildMiner} to extract this information from the build files. In an industrial case study, alternative implementations are handled by a configuration management tool rather than by \texttt{MAKE} scripts. In this case, we had to develop a new build model extractor to collect this information. However, the existing data model did not require any modifications. Also other extractors and analysis plug-ins could be reused without any modifications on Linux as well on the industrial case study.

\subsection{Extension for New Experiments}
\label{sec:Extension}
KernelHaven can be configured by a configuration file to be able to fit to the specific needs of the desired experiment (cf.\ Section~\ref{sec:Concepts}). In addition, KernelHaven offers extension points to allow the integration of extractor and analysis plug-ins, if simple configuration is not sufficient to integrate a new concept.
These extractors and analyses 
use the common data representation (cf. Section~\ref{sec:Data}) and may even reuse existing plug-ins as part of their process. Further, extractors may extend the elements of the data models to provide custom information. As the 
implementation of a new extractor or analysis is only a matter of defining the desired parent class, e.g., the abstract code extractor class for a new code extractor, this section focuses on the (re-)use of the data models and existing plug-ins.

The data models of each pipeline represent interfaces between extractors and analyses. Hence, each new extractor needs to return (some of the) respective model elements, which in turn are provided to the requesting analysis. For example, the Undertaker code extractor creates and returns \texttt{CodeBlock} elements as illustrated in Figure~\ref{fig:Code Model}. However, for some experiments this element may not be sufficient to extract all relevant data from code files, while the \texttt{SyntaxElement} on the other hand is too expressive. In this situation, the new extractor can introduce its own element by defining a new class derived from \texttt{CodeBlock} of the common data model. Due to inheritance, the infrastructure transparently provides this new code block to an analysis. This extension mechanism can in principle be applied by each type of extractor.

Analysis plug-ins may reuse other analyses as part of their implementation. For instance, the feature effect analysis described in Section~\ref{sec:Concepts} actually consists of a set of analyses, each with its individual purpose. First, we developed an analysis plug-in to discover all presence conditions (code and build) in a product line. Second, another analysis takes these conditions as an input to calculate \textit{feature effect} constraints \cite{NadiBergerKastner+15}. In this way, each analysis plug-in becomes reusable and can be combined to a pipeline of (atomic) analysis plug-ins.

KernelHaven offers two alternatives to realize such analysis pipelines. Figure~\ref{fig:Analysis pipeline} shows an example of both alternatives for the realization of the feature effect analysis.
\begin{enumerate}
	\item The wiring of the analysis plug-ins may be done in code, if the previous analysis plug-ins need not to be exchangeable. In this case, researchers need to create a new class which inherits from \texttt{PipelineAnalysis} and overrides the \texttt{createPipeline}-method as shown in the left listing. This method specifies the wiring of the analysis plug-ins, but also the required extractor pipelines. As a consequence, the user needs only to specify this class in the configuration scripts in order to run all analysis plug-ins and the required extractors to retrieve the desired output.
  \item KernelHaven also offers the possibility to configure the complete pipeline in a configuration file as shown in the right listing. For this, the user has to select the \texttt{ConfiguredPipelineAnalysis} as done in Line~1 and to model the pipeline with an integrated DSL as done in Lines~2--5. This alternative allows to flexibly exchange partial analysis steps without 
  writing new code.
\end{enumerate}

In both cases, it is possible to specify for each analysis the required input. In the example of Figure~\ref{fig:Analysis pipeline}, the detection of presence conditions requires the output of a build model (\texttt{bmComponent}) and a code (\texttt{cmComponent}) extractor. The analysis of feature effects requires only the results of the previous analysis, however, both approaches allow to specify further input sources if required.

\subsection{Support for Reproduction}
\label{sec:Archiving}
Reproduction of experiments in research and, in particular, in computer science becomes a major problem today \cite{EichelbergerSassSchmid16, CitoGall16, Boettiger15}. KernelHaven addresses these problems through the following concepts:

\textbf{Intermediate Results.} KernelHaven offers the optional possibility to save intermediate results in human readable form. This includes the cached information of the extractors (cf.\ Section~\ref{sec:Caching}), intermediate analysis results, and the final results of the actual analysis. This facilitates a manual verification of each processing step in the sense of a black-box test.

An example of how to obtain such intermediate results is shown in the right part of Figure~\ref{fig:Analysis pipeline}. KernelHaven stores only the results of the last processing step by default. In this example, this would only be the results of the feature effect analysis. The optional parameter in Line~6 allows to specify any number of intermediate analysis results, which should also be saved. Further, Line~7 specifies Excel sheets as output format (\texttt{ExcelBook}).  In the example, the results of the presence conditions will be written to a separate sheet in the resulting Excel document. Alternatively it would be possible to save the results as separate CSV-files or another output format, which may be defined by a plug-in.

\textbf{Dependency-free Plug-ins.} In the previous sections we already described the integration of external tools by means of the plug-in architecture of KernelHaven. We designed these plug-ins to integrate all necessary external tools and dependent libraries as far as possible to avoid a dependency hell. As a result, KernelHaven and most of its plug-ins require only Java. However, some of the plug-ins are compiled for a specific operating system only.

For instance, the Undertaker-based extractor requires Linux to be able to execute the included binary. However, it does not require the installation of any further libraries, because all required libraries are linked statically.

\textbf{Documentation of Experiments.} KernelHaven supports automatic archiving of all relevant plug-ins, input, output, and intermediate data to reproduce an experiment. In particular, KernelHaven takes care to bundle itself as well as all available plug-ins and the used configuration file, which specifies the experimental setup, i.e., which plug-ins are used and how the data is passed from the extractors through the different analysis steps. Further, the archive includes the input data and any resulting information, e.g., cached extraction data, log files, intermediate, and final analysis results.

The resulting archive contains all relevant information for an external review. Further, it supports reproduction of an entire experiment on a new machine. This only requires the installation of Java on an appropriate operating system.

\subsection{Parallelization}
\label{sec:Parallelization}
KernelHaven supports parallel data extraction and analysis on two different levels: i) each extraction pipeline in Figure~\ref{fig:Architecture} runs as an individual thread and ii) extractors as well as analyses may run concurrently. This parallelization enables an analysis to immediately start its evaluation as soon as sufficient data is available. In particular, an analysis plug-in may start even while a required analysis plug-in, which serves as input, is not completely finished.

The main advantage of this feature is that parallelization is handled transparently by the infrastructure. Hence, researchers do not need to care about it and can focus on the implementation of their approach. For instance, code extractors are designed for the translation of a single file. The infrastructure creates multiple instances of them to translate several code files in parallel. This mechanism is also available for analyses. Further, the infrastructure manages the entire lifecycle of each thread, such that researchers can focus on their core algorithms instead of spending effort on parallelization.

In some situations, however, parallelization may not be desired as the respective extractor or analysis requires all information to be available before being able to produce their results. For example, the TypeChef code extractor may use the parsed variability model to exclude irrelevant parts from the variability aware AST. While this is in general a design decision, individual configuration parameters can be used to switch from parallel to non-parallel processing for each extractor and analysis. In this way, the same extractor or analysis can be used in both a parallel and a non-parallel setup.

\subsection{Data Caching}
\label{sec:Caching}
The core infrastructure of KernelHaven offers a mechanism to save and reuse extracted information from code, build, and variability model files. This data caching feature is implicitly available for each extractor and can be enabled or disabled by two distinct configuration options in a configuration file: one for writing the extracted information to the hard disc and one for reading the cached data instead of executing the extractors. As these two options are available for each pipeline (extractor), either one or multiple pipelines may cache and reuse their data. The main benefit of this feature is that conducting the same experiment multiple times as well as conducting similar experiments (same extracted information but different analysis) takes significantly less time than processing the inputs again for every experiment.

The model elements created and returned by an extractor define the data to be cached. For example, the TypeChef extractor creates a set of \texttt{SyntaxElement}s (cf. Figure~\ref{fig:Code Model}), which the infrastructure passes to an analysis. If the configuration option for caching the extracted code information is enabled, the infrastructure also serializes these code model elements to the hard disc. The next experiment requiring the same information may enable the configuration option for reading this cached information. In that case, the infrastructure reads the serialized code model elements from hard disc instead of executing the code extractor leading to significant performance improvements.

\section{Conclusion}
\label{sec:Conclusion}
We presented KernelHaven as an experimentation workbench to support and simplify experiments in the domain of variability-based static analysis of C-preprocessor Software Product Lines (SPL). This novel methodology aims at providing a reusable infrastructure along with some standard capabilities, which researchers can (re-)use for their experiments. KernelHaven realizes this methodology, e.g., to evaluate approaches for correctness checks or metrics, for its specific domain. It was already applied in two different setups to perform the feature effect analysis on Linux, which is one of the largest SPL known in research, as well as on an industrial SPL of similar size and complexity. Further, the general concepts presented here may be adopted by researchers in other software engineering domains. Hence, we expect two types of follow-ups in the future: the integration of additional analyses as KernelHaven is more established and the transfer of concepts to other domains to support experimental research.

\vspace*{-2ex}
\begin{acks}
This work is partially supported by the ITEA3 project $\text{REVaMP}^2$, funded by the \grantsponsor{01IS16042H}{BMBF (German Ministry of Research and Education)}{https://www.bmbf.de/} under grant \grantnum{01IS16042H}{01IS16042H}. Any opinions expressed herein are solely by the authors and not by the BMBF.
  
We would like to thank the following contributors: Moritz Fl\"oter, Adam Krafczyk, Alice Schwarz, Kevin Stahr, Johannes Ude, Manuel Nedde, Malek Boukhari, and Marvin Forstreuter.
\end{acks}

\begin{spacing}{0.95}
\bibliographystyle{ACM-Reference-Format}
\bibliography{literature}
\end{spacing}

\end{document}